\documentclass[twocolumn,prb,showpacs,preprintnumbers,amsmath,amssymb]{revtex4-1}
\usepackage{graphicx}% Include figure files
\usepackage{dcolumn}% Align table columns on decimal point
\usepackage{bm}% bold math
\usepackage{verbatim}

\begin{document}

%\preprint{APS/123-QED}

\title{Magnetization vector in the reversible region of a highly anisotropic cuprate superconductor: anisotropy factor and the role of 2D vortex fluctuations}

\author{Jes\'us Mosqueira}
\author{Ram\'on I. Rey}%
\author{F\'elix Vidal}
\affiliation{LBTS, Departamento de F\'isica da Materia Condensada, Universidade de Santiago de Compostela, E-15782 Santiago de Compostela, Spain}

\date{\today}% It is always \today, today,

\begin{abstract}
By using a high quality Tl$_2$Ba$_2$Ca$_2$Cu$_3$O$_{10}$ (Tl-2223) single crystal as an example, the magnetization vector was probed in the reversible region of highly anisotropic cuprate superconductors. For that, we have measured its components along and transverse to the applied magnetic field for different crystal orientations. The analysis shows that the angular dependence of the \textit{perpendicular} component of the magnetization vector follows the one predicted by a London-like approach which includes a contribution associated with the thermal fluctuations of the 2D vortex positions. For the Tl-2223 crystal studied here, a lower bound for the anisotropy factor was estimated to be about 190.
\end{abstract}

\maketitle

\section{Introduction}

Due to their weakly coupled layered structure, most of the high-T$_c$ cuprate superconductors (HTSC) are highly anisotropic, a feature which affects many of the most important properties of these materials, including their response to magnetic and electric fields.\cite{tinkham} The central parameter when describing this anisotropic behaviour is the so-called anisotropy factor, $\gamma$, which may be defined as  $\gamma=(m^*_c/m^*_{ab})^{1/2}$, where $m^*_{ab}$ and $m^*_c$ are the effective supercarrier masses for currents flowing in the $ab$ (CuO$_2$) layers and, respectively, along the $c$ axis.\cite{tinkham} Particularly useful in describing the phenomenology of the HTSC are the relationships between $\gamma$ and the two dimensionless scales defined by the ratio between the superconducting layers periodicity length, $s$, and the two superconducting characteristic lengths, the coherence length in the $ab$ planes, $\xi_{ab}$, and the penetration length for a magnetic field applied perpendicularly to these planes, $\lambda_{ab}$.\cite{tinkham,feinberg94} 
On the one side, $\xi_{ab}/s$ represents the anisotropy above which the material behaves as a stack of Josephson-coupled superconducting layers, and a Lawrence-Doniach (LD) description may be required.\cite{LD} On the other, $\gamma>\lambda_{ab}/s$ is the threshold for two-dimensional (2D) effects to appear in the vortex structure: for instance, a \textit{kinked} vortex lattice, \cite{feinberg90,feinberg92} a decoupled set of vortex lattices parallel and perpendicular to the \textit{ab} layers,\cite{kes90,blkprb92} or the vortex \textit{lock-in} onto the \textit{ab} layers direction,\cite{theodorakis90,feinberg90lockin1,feinberg90lockin2,steinmeyer94,vulcanescu94,janossy95} may appear in tilted crystals when $\tan\theta\stackrel{>}{_\sim}\gamma$, where $\theta$ is the angle between the applied magnetic field and the crystal's $c$ axis.
The interest of a precise knowledge of the anisotropy factors of the HTSC is also enhanced by the present debate on the nature of the Meissner transition in these cuprate superconductors, mainly in the subdoped regime, and on the applicability of the different Ginzburg-Landau like approaches to describe at a phenomenological level their behavior.\cite{timusk,mosqueira09}

In spite of its relevance, the anisotropy factors of some of the most important HTSC families, mainly in the case of the most anisotropic ones based in Bi, Tl, or Hg, are still not well settled, the dispersion in the different proposed values being very large. For instance, in the case of Tl-2223 $\gamma$ values as low as 2.8 and up to 900 were proposed.\cite{steinmeyer94,gammatl2223,maignan94} 
One reason for such a dispersion lies in the samples' structural and stoichiometric quality: for instance, a $\Delta \alpha\sim1^\circ$ mosaic spread would strongly affect the determination of the anisotropy factor if it is larger than $\gamma\sim1/\tan\Delta\alpha\sim50$. Also, $\gamma$ may be strongly dependent on the doping level, and doping-level variations from sample to sample may lead to large differences in the observed $\gamma$ value.
Another reason lies in the procedures currently used to determine $\gamma$. For instance measurements of the ratio between the \textit{longitudinal} and \textit{transverse} magnetizations or resistivities require aligning the crystal with magnetic or electric fields with accuracies of the order of $1/\arctan(\gamma^{-1})$, i.e., close to $\sim0.5^\circ$ when $\gamma=100$.
These difficulties associated with the crystal alignment may be overcome by measuring the $\theta$ dependence of the magnetic torque per unit volume, $\vec\tau=\mu_0\vec M\times \vec H$, in the reversible mixed state: The 3D anisotropic Ginzburg-Landau approach (3D-aGL) predicts a strong dependence of $\tau$ on $\gamma$ mainly close to $\theta=\pi/2$.\cite{kogan88a,hao92,buzdin94} Therefore, a fine angular sampling around this $\theta$ value will in principle suffice for a precise $\gamma$ determination. However, $\vec\tau$ presents a contribution from the components of the magnetization in the directions parallel, $M_\parallel$, and perpendicular, $M_\perp$, to the \textit{ab}-layers, $|\vec\tau|=\mu_0H(M_\perp \sin\theta-M_\parallel \cos\theta)$, and this last component may be strongly affected, mainly in highly anisotropic superconductors, by thermal fluctuations of the 2D vortices (pancakes).\cite{blk92,kogan93} This could limit the applicability of the 3D-aGL approach to extract $\gamma$ from these measurements.\cite{martinez92}

In this paper, we present measurements of the angular dependence of the magnetization vector, $\vec M$, in the reversible mixed state in a high quality single crystal of the strongly anisotropic Tl$_2$Ba$_2$Ca$_2$Cu$_3$O$_{10}$ (Tl-2223). For that, we measured simultaneously the longitudinal, $M_L$ and the transverse, $M_T$, magnetization components relative to the applied magnetic field. This procedure allows to overcome the difficulties stressed above: these data provides a stringent check of the applicability of the 3D-aGL theory, which predicts that the ratio $M_T/M_L$ depends only on $\gamma$ and $\theta$.\cite{kogan88} In addition, one may obtain the perpendicular and parallel components of $\vec M$ ($M_\perp$ and $M_\parallel$) and then to check the relevance of the superconducting fluctuation effects on $M_\perp$. Let us note already here that until now only very few works include simultaneous measurements of $M_L$ and $M_T$ in the reversible mixed state, and they correspond to samples with a relatively moderate anisotropy.\cite{tuominen90,pugnat95,bruckental06} 
The analyses of our present data shows that the angular dependence of $M_\perp$ follows the one predicted by a London approach for layered superconductors which includes the contribution of the pancakes fluctuations. The breakdown of this approach observed when $\theta$ is only 0.3$^\circ$ away from $\pi/2$ is attributed to the mosaic spread present in the sample, and allowed to estimate a lower bound for the anisotropy factor of Tl-2223 as high as $\sim$190.

\section{Experimental details and results}

The Tl-2223 sample used in this work is a plate-like single crystal (1.1$\times$0.75$\times$0.226 mm$^3$) with the $c$ crystallographic axis perpendicular to the largest face. Details of its growth procedure and characterization by Maignan \textit{et al.} may be found in Ref.~\onlinecite{maignan94}. It is the same crystal used in the magnetization measurements with $H\perp c$ presented in Ref.~\onlinecite{mosqueira07}, which presents a sharp low-field diamagnetic transition ($T_c=122\pm1$ K) and a large Meissner fraction ($\sim$80\%). Of a great importance for the present work is the excellent alignment of the crystal $c$-axis: the rocking curves of (00$\ell$) reflections (inset in Fig.~1) may be fitted to a Lorentz distribution of only $\sim0.08^\circ$ wide (FWHM), which will allow to study the angular dependence of the magnetization vector in almost the entire angular range.

The magnetization measurements were performed with a commercial SQUID magnetometer (Quantum Design, model MPMS), equipped with detectors for the components of the magnetic moment along ($m_L$) and transverse ($m_T$) to the applied magnetic field. The sample was glued with a minute of GE varnish to a rotating sample holder (also from Quantum Design) which allowed rotations about the $L$ and $T$ axes. In both cases, the orientation may be specified within 0.1$^\circ$, with a reproducibility of $\pm0.5^\circ$ about the $L$ axis and $\pm1^\circ$ about the $T$ axis. 

\begin{figure}[t]
\includegraphics[width=\columnwidth]{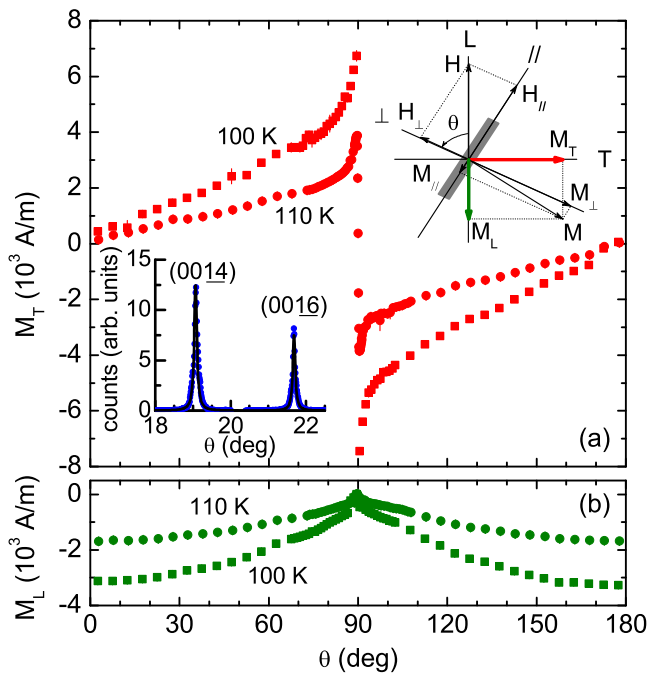}
\caption{(Color online) Angular dependence of the transverse (a) and longitudinal (b) components of the magnetization vector. The diagram shows the crystal orientation with respect to the longitudinal-transverse ($L,T$) frame and to the perpendicular-parallel ($\perp,\parallel$) frame (the crystal $c$ axis is along the $\perp$ axis). Lower inset: rocking curves of the (00\underline{14}) and (00\underline{16}) reflections. They reveal the presence of a $c$-axis mosaic spread following a Lorentz distribution of only $\sim$0.08$^\circ$ wide (FWHM).}
\end{figure}

The $\theta$ dependence of $m_L$ and $m_T$ was measured at constant temperatures below $T_c$ (100 K and 110 K) under a 1~T magnetic field. These values were chosen because they are well inside the reversible region of the $H-T$ phase diagram even for $\theta=90^\circ$, and in the low magnetic field regime [$H\ll H_{c2}(\theta)$] even for $\theta=0^\circ$, both facts allowing an easy comparison with the theory (see below). The resulting longitudinal and transverse magnetizations ($M_{L,T}=\rho m_{L,T}/M$, where $M=1.3$~mg is the crystal mass and $\rho=6.96$~g/cm$^3$ is the \textit{theoretical} density) are presented in Fig.~1, where the choice for the ($L$,$T$) and ($\perp$,$\parallel$) frames is also shown. $M_L$ in that figure is already corrected for a small contribution to the magnetic moment coming mainly from the sample holder. This contribution (of about $-7\times10^{-8}$~Am$^2$, and $\theta$-independent within $\sim3$\%) was determined at a temperature well above $T_c$ (200~K) where the effect of superconducting fluctuations is negligible, but taking into account that it presents a slight temperature dependence (about -5$\times10^{-11}$~Am$^2$ per K). The uncertainty in the corrected $M_L$ at 110~K is estimated to be less than $\pm 10$\%. 

As expected, for $\theta=0^\circ$ or $\theta=90^\circ$ $M_T$ vanishes. This discards the presence of any spurious contribution to the measured $M_T$ coming from the longitudinal component, $M_L$. For an arbitrary $\theta$ value $M_T$ it is not null because of the crystal anisotropy. In view of the large $|M_T|$ values (for \mbox{$|\theta-90^\circ|\stackrel{<}{_\sim}30^\circ$} it is even larger than $|M_L|$) it may be already inferred that the crystal is strongly anisotropic. Close to $\theta=90^\circ$, $M_L\approx0$ and $|M_T|$ presents a sharp peak which was not observed in the pioneering  magnetization vector experiments by Tuominen \textit{et al.} in Bi$_2$Sr$_2$CaCu$_2$O$_8$ (Bi-2212),\cite{tuominen90} for which an anisotropy factor of $\gamma=16.7$ was obtained. As we will see below, this already suggests that the anisotropy factor in Tl-2223 must be much larger than this value.

\section{data analysis}

In the framework of the 3D-aGL theory, in the London region [i.e., for $H_{c1}(\theta)\ll H\ll H_{c2}(\theta)$, where $H_{c1}$ and $H_{c2}$ are the lower and upper critical fields, respectively], $M_L$ and $M_T$ are given by\cite{kogan88}
\begin{equation}
M_L=-M_0\varepsilon(\theta)\ln\left[\frac{\eta H_{c2}^\perp}{H\varepsilon(\theta)}\right]
\end{equation}
and
\begin{equation}
M_T=M_0(1-\gamma^{-2})\frac{\cos\theta\sin\theta}{\varepsilon(\theta)}\ln\left[\frac{\eta H_{c2}^\perp}{H\varepsilon(\theta)}\right].
\end{equation}
Here $H_{c2}^\perp\equiv H_{c2}(\theta=0)$, $M_0\equiv\phi_0/8\pi\mu_0\lambda_{ab}^2$, $\varepsilon(\theta)=(\cos^2\theta+\sin^2\theta/\gamma^2)^{1/2}$, $\eta$ a constant about the unity related to the vortex lattice structure, $\phi_o$ the magnetic flux quantum, and $\mu_0$ the magnetic permeability. It has been early noted \cite{kogan88} that the ratio $M_T/M_L$ is only dependent on $\theta$ and $\gamma$:
\begin{equation}
\frac{M_T}{M_L}=(1-\gamma^2)\frac{\sin\theta\cos\theta}{\sin^2\theta+\gamma^2\cos^2\theta}.
\label{ratio}
\end{equation}
This fact has been used in previous works to determine the anisotropy factor from simultaneous measurements of $M_L(\theta)$ and $M_T(\theta)$.\cite{tuominen90,pugnat95,bruckental06} In Fig.~2(a) we present the $\theta$ dependence of $M_T/M_L$, as results from the data in Fig.~1. For comparison we have also included the data from Ref.~\onlinecite{tuominen90} for Bi-2212, which is also highly anisotropic. The lines in this figure correspond to Eq.~(\ref{ratio}), evaluated with several $\gamma$ values. In the case of Bi-2212, as shown in Ref.~\onlinecite{tuominen90}, $\gamma=16.7$ leads to a good agreement in the whole $\theta$ range. In the case of Tl-2223 the agreement is very good for $|\theta-90^\circ|\stackrel{>}{_\sim}5^\circ$, independently of the $\gamma$ value used above $\sim30$. However, as shown in Fig.~2(b), closer to $\theta=90^\circ$ where the theoretical $M_T/M_L$ is strongly dependent on $\gamma$, the disagreement is very important: by using $\gamma=30$, the amplitude of the peak observed close to $\theta=90^\circ$ is underestimated. On the contrary, if a larger value is used, i.e., $\gamma=50$, the peak amplitude is reproduced but the fit to the data for $|\theta-90^\circ|\stackrel{<}{_\sim}5^\circ$ is poor. 
Such a disagreement is not significantly improved by using a generalization of the 3D-aGL approach to quasi-2D layered superconductors proposed by Hao\cite{hao}. This author suggested that the $\theta$ dependence of the upper critical field may be parametrized by
\begin{equation}
\left[\frac{H_{c2}(\theta)\cos\theta}{H_{c2}^{\perp}}\right]^n+\left[\frac{H_{c2}(\theta)\sin\theta}{H_{c2}^{\parallel}}\right]^2=1, 
\label{hc2hao}
\end{equation}
with $1\leq n\leq2$. If $n=2$, the above equation reduces to the result of the effective mass model $H_{c2}(\theta)=H_{c2}^\perp/\varepsilon(\theta)$, while if $n=1$ it leads to the result of Tinkham for a thin film. In the framework of this model it is found
\begin{equation}  
\frac{M_T}{M_L}=\frac{2g_\theta^{n-2}\cos\theta\sin\theta-n\gamma^2\cos^{n-1}\theta\sin\theta}{n\gamma^2\cos^n\theta+2g_\theta^{n-2}\sin^2\theta},
\label{ratiohao}
\end{equation}
where $g_\theta\equiv H_{c2}^\perp/H_{c2}(\theta)$ as results from Eq.~(\ref{hc2hao}). A comparison of Eq.~(\ref{ratiohao}) (evaluated in the extreme case that $n=1$) with the experimental data is presented in Fig.~2(c). As may be clearly seen, for $\gamma$ values in the range 30$-$50 Eq.~(\ref{ratiohao}) reproduced the qualitative features of $M_T/M_L$ around $\theta=90^\circ$ but, again, a quantitative agreement occurs only $\sim5^\circ$ away from $\theta=90^\circ$. 

\begin{figure}[t]
\includegraphics[width=\columnwidth]{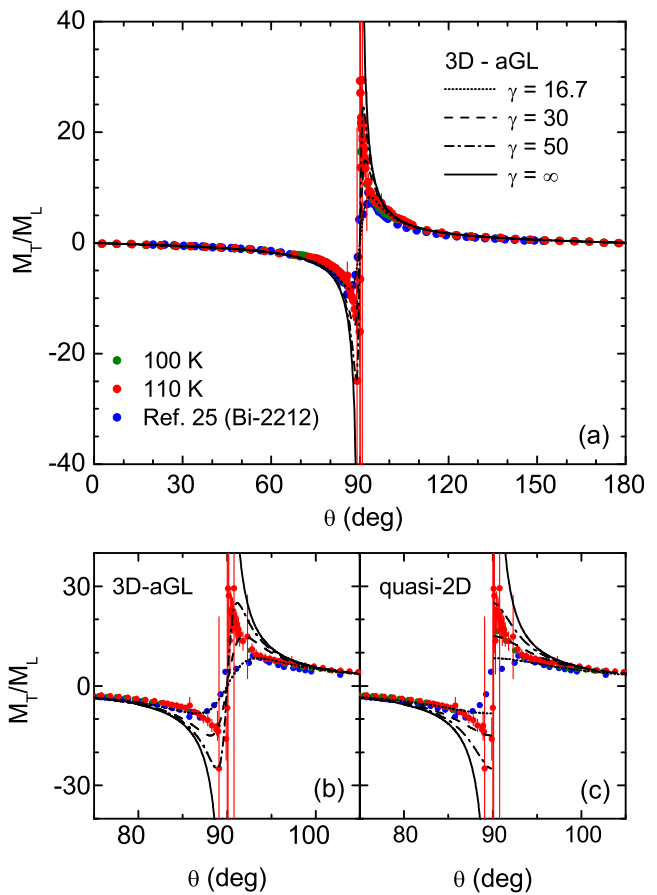}
\caption{(Color online) (a) Angular dependence of the $M_T/M_L$ ratio (data taken at 100 K are completely overlapped by the ones taken at 110 K). For comparison, data from a Bi-2212 crystal taken from Ref.~\onlinecite{tuominen90} are also included. The lines correspond to Eq.~(\ref{ratio}) evaluated with different $\gamma$ values. (b) Detail around $\theta=90^\circ$. (c) Comparison of the data around $\theta=90^\circ$ with the generalization of the 3D-aGL approach to quasi-2D superconductors developed in Ref.~\onlinecite{hao} [Eq.~(\ref{ratiohao})]. In all figures, the vertical lines are the experimental error bars.}
\label{figratio}
\end{figure}

Among the possible causes for the observed discrepancies are: i) from the experimental point of view, the analysis of $M_T/M_L$ for $\theta\sim90^\circ$ in highly anisotropic materials is complicated by the fact that $M_L(90^\circ)\propto\gamma^{-1}$, which tends to zero for large $\gamma$ values. In this way, the experimental uncertainty affecting $M_L$ (mainly coming from the sample holder contribution) makes the region close to $90^\circ$ experimentally unaccessible (note the large experimental bars in the $M_T/M_L$ data points close to $\theta=90^\circ$). ii) From the theoretical point of view, the effect of thermal fluctuations on the 2D vortices positions was not taken into account on deriving Eqs.~(\ref{ratio}) and (\ref{ratiohao}).\cite{blk92} However, it was later shown that these fluctuations may be very important for temperatures close to $T_c$ when $H\perp ab$.\cite{kogan93} In particular, they give rise to the so-called \textit{crossing point} in which the magnetization at a temperature few Kelvin below $T_c$ is independent of the applied magnetic field.\cite{kes91} 
While fluctuations of \textit{pancake} vortices are reasonably well
established, there is no experimental evidence or theoretical basis
for corresponding positional fluctuations of Josephson vortices in the
case with $H\parallel ab$; consequently, it is not expected that the ratio $M_T/M_L$ should follow Eq.~(\ref{ratio}) or its generalization, Eq.~(\ref{ratiohao}).

\begin{figure}[t]
\includegraphics[width=\columnwidth]{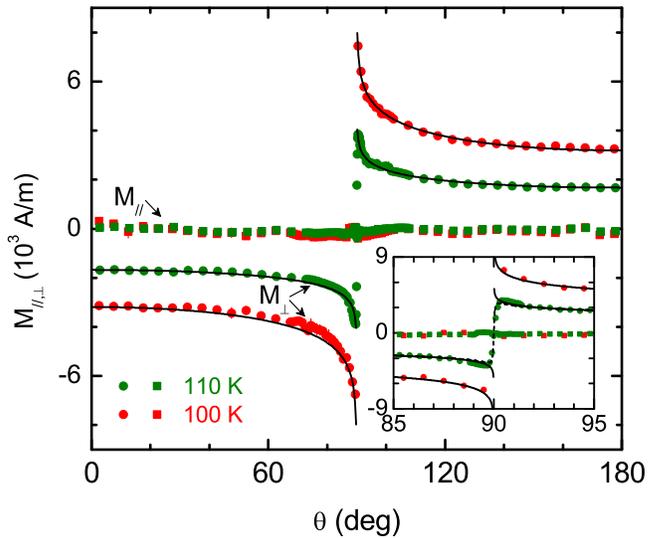}
\caption{(Color online) Angular dependence of the perpendicular and parallel components of the magnetization vector. The lines correspond to Eq.~(\ref{blktheta2}) evaluated with the Tl-2223 parameters obtained in Ref.~\onlinecite{mosqueira07} from $M_\perp(T,H)$ measurements with $H\perp ab$ (solid lines). Inset: detail around $\theta=90^\circ$. The dashed line in the 110 K measurement corresponds to Eq.~(10), which takes into account the presence of a narrow $c$-axis mosaic spread (see main text for details).}
\label{perppara}
\end{figure}

An alternative way of analyzing $M_L(\theta)$ and $M_T(\theta)$ consist of obtaining the $\vec M$ components in the directions perpendicular $M_\perp$ and parallel $M_\parallel$ to the $ab$ layers through the linear transformation
\begin{equation}
\left(
\begin{array}{c}
M_{\perp}\\
M_\parallel
\end{array}
\right)
=
\left(
\begin{array}{cc}
\cos\theta & -\sin\theta \\
\sin\theta & \cos\theta  
\end{array}
\right)
\left(
\begin{array}{c}
M_L\\
M_T
\end{array}
\right).
\label{matrix}
\end{equation}
The result is presented in Fig.~3. As expected, a first advantage of this procedure is that in the vicinity of $\theta=90^\circ$, $M_\perp$ is much less affected than $M_T/M_L$ by the experimental uncertainties in $M_L$. In addition, the data for $M_\perp$ may be directly compared with the Bulaevskii, Ledvig and Kogan (BLK) expression which takes into account the above mentioned thermal fluctuations of 2D vortices,\cite{blk92}
\begin{eqnarray}
M_\perp^{\rm BLK}=-M_0\ln\left(\frac{\eta H_{c2}^\perp}{H}\right)
+M_1\ln\left(\frac{M_1}{M_0c_1}\frac{\eta H_{c2}^\perp}{H}\right),
\label{blk}
\end{eqnarray}
where $M_1=k_BT/\phi_0s$, and $c_1$ is a constant of the order of the unity. The first term corresponds to the conventional London contribution, and the second is the one due to thermal fluctuations. As Eq.~(\ref{blk}) follows from the Lawrence-Doniach model, the scaling transformation relating the properties of an arbitrarily oriented 3D anisotropic superconductor with the ones in a isotropic superconductor,\cite{klemm80,blatter92,hao92} cannot be used to obtain the $\theta$ dependence of $M_\perp^{\rm BLK}$. However, for an extremely anisotropic compound one may assume that the perpendicular component of the magnetization is only dependent on the perpendicular component of the magnetic field, and approximate
\begin{equation}
M_\perp^{\rm BLK}(T,H,\theta)\approx \pm M_\perp^{\rm BLK}(T,H|\cos\theta|,0),
\label{blktheta}
\end{equation}
where the + sign applies to $0\leq\theta<90^\circ$ and the $-$ to \mbox{$90^\circ<\theta\leq180^\circ$}.
The angular dependence of $M_\perp^{\rm BLK}$ may be then expressed as
\begin{equation}
M_\perp^{\rm BLK}(\theta)=\pm M_\perp^{\rm BLK}(0)\pm(M_0-M_1)\ln|\cos\theta|,
\label{blktheta2}
\end{equation}
with the same convention for the signs. $M_\perp^{\rm BLK}(0)$ may be determined directly from the data in Fig.~3. Also, from Eq.~(\ref{blk}) it follows that $M_0-M_1=dM_\perp^{\rm BLK}(H,0)/d\ln H$. This last quantity has been determined in detail in the same crystal \cite{mosqueira07} and, as predicted by Eq.~(\ref{blk}), it is a constant for each temperature provided that the sample is in the low field region, $H\ll H_{c2}^\perp(T)$. We have from Ref.~\onlinecite{mosqueira07} that $M_0-M_1=831$ A/m at $T$=100 K and 412 A/m at $T$=110 K. This allows to compare Eq.~(\ref{blktheta2}) with the experimental data \textit{without free parameters}. As shown in Fig.~\ref{perppara}, the agreement is excellent in almost the whole angular range, \cite{unpub} up to $|\theta-90^\circ|\leq\alpha_0\approx0.3^\circ$. 
A possible cause for the disagreement closer to $\theta=90^\circ$ remains in the slight mosaic spread present in the sample. To confirm this, we roughly estimated the effective perpendicular magnetization through the angular average

\begin{equation}
\langle M_\perp^{\rm BLK}(\theta)\rangle=\int_{-\pi/2}^{\pi/2}\delta(\alpha)M_\perp^{\rm BLK}(\theta-\alpha)\cos\alpha\;d\alpha,
\label{mosaic}
\end{equation}
where $\delta(\alpha)$ is the volume distribution of domains with the $c$ axis tilted an angle $\alpha$ with respect to the average $c$ axis, which may be approximated by a Lorentzian distribution with 0.08$^\circ$ FWHM (see above). As it is shown in inset of Fig.~3, Eq.~(\ref{mosaic}) (dashed line) accounts for the $M_\perp(\theta)$ \textit{rounding} observed when $|\theta-90^\circ|\leq\alpha_0$. The possible 2D effects on the vortex lattice, which are expected to occur when $\tan\theta>\gamma$, may be hidden by such a rounding, making difficult a precise determination of $\gamma$. However, a lower bound may be estimated from the onset of the $M_\perp(\theta)$ rounding, leading to a value as high as $\gamma_{\rm min}\approx\tan(90^\circ-\alpha_0)\approx190$.

Let us finally mention that the London approach alone (without the term associated to vortex fluctuations) leads to an angular dependence for $M_\perp$ similar to the one of Eq.~(\ref{blktheta2}). This, together with the fact that in highly anisotropic compounds $M_\parallel\approx0$, explains why the angular dependence of the magnetic torque may be successfully described in a wide $\theta$-region by the London approach, without taking into account the contribution of vortex fluctuations. However, as early shown in Ref.~\onlinecite{kogan93} (and in Ref.~\onlinecite{mosqueira07} for Tl-2223), the application of the London theory to highly anisotropic HTSC leads to unphysical values for the superconducting parameters, mainly close to the so-called \textit{crossing point} temperature where $M_\perp$ is independent of the applied magnetic field.

\section{Summary}

The magnetization vector of a highly anisotropic cuprate superconductor, a high quality Tl$_2$Ba$_2$Ca$_2$Cu$_3$O$_{10}$ single crystal, has been studied in the reversible region of the $H$-$T$ phase diagram. For that, we have measured simultaneously the angular dependence of its longitudinal ($M_L$) and transverse ($M_T$) components in the London region. Then, these data have been compared with the 3D-aGL theory, in particular through the relationship $M_T/M_L(\theta)$, which on the grounds of this approach depends only on the anisotropy factor. 
This procedure, which was successfully used in the literature to probe the applicability of the 3D-aGL approach and to obtain $\gamma$ in moderately anisotropic samples, was shown to be inappropriate in the present case: on the one side, there is a large experimental uncertainty associated to the vanishing of $M_L$ when $\theta\to90^\circ$ due to the high anisotropy of Tl-2223. On the other, $M_L$ and $M_T$ may be affected in a different way by thermal fluctuations of 2D vortices (which were shown to affect strongly the magnetization perpendicular to the ab layers) thus limiting the applicability of the 3D-aGL approach. As an alternative, we transformed $M_L$ and $M_T$ into the components perpendicular ($M_\perp$) and parallel ($M_\parallel$) to the $ab$ layers. The analysis of $M_\perp(\theta)$ in terms of the BLK model which takes into account the above mentioned fluctuation effects, leads to an excellent agreement in the whole $\theta$ range, except in a narrow interval ($0.6^\circ$ wide) around $\theta=90^\circ$. This breakdown was correlated with the mosaic spread present in the sample, and set a lower bound for the anisotropy factor in this compound of $\gamma\approx190$.

\acknowledgements

This work was supported by the Spanish Ministerio de Educaci\'on y Ciencia (Grant No. FIS2007-63709), and the Xunta de Galicia (Grant No. 07TMT007304PR). We acknowledge A. Maignan and A. Wahl for providing us the sample, and J. Ponte his valuable help with the rotating sample holder.


\begin{references}

\bibitem{tinkham}See, e.g., M. Tinkham, in \textit{Introduction to Superconductivity} (McGraw-Hill, New York, 1996), Chap. 9.

\bibitem{feinberg94}For reviews on the effects in the magnetization associated with the anisotropy of the HTSC and earlier references see, e.g., D. Feinberg, J. Phys. III France {\bf4}, 169 (1994); G. Blatter, M. Feigel'man, V.B. Geshkenbein, A.I. Larkin, and V.M. Vinokur, Rev. Mod. Phys. {\bf66}, 1125 (1994); E.H. Brandt, Rep. Prog. Phys. {\bf58}, 1465 (1995).

\bibitem{LD} Lawrence W.E., and Doniach S., Proc. of the 12th Conf. on Low-Temperature Physics, Ed. by  E. Kanda (Keigaku, 1970), p. 361.  
  
\bibitem{feinberg90}D. Feinberg and C. Villard, Phys. Rev. Lett. {\bf65}, 919 (1990).

\bibitem{feinberg92}D. Feinberg, Physica C {\bf194}, 126 (1992).

\bibitem{blkprb92}L.N. Bulaevskii, M. Ledvij, and V.G. Kogan, Phys. Rev. B {\bf46}, 366 (1992).	
	
\bibitem{kes90}P.H.	Kes, J. Aarts, V.M. Vinokur, and C.J. van der Beek, Phys. Rev. Lett. {\bf64}, 1063 (1990).

\bibitem{theodorakis90}S. Theodorakis, Phys. Rev. B {\bf42}, 10172 (1990).

\bibitem{feinberg90lockin1}D. Feinberg and C. Villard, Mod. Phys. Lett. B {\bf4}, 9 (1990).	

\bibitem{feinberg90lockin2}For a review, see e.g., D. Feinberg and A.M. Ettouhami, Int. J. Mod. Phys. B {\bf7}, 2085 (1993).	
	
\bibitem{steinmeyer94}F. Steinmeyer, R. Kleiner, P. M\"{u}ller, H. M\"{u}ller, and K. Winzer, Europhys. Lett. {\bf25}, 459 (1994).

\bibitem{vulcanescu94}V. Vulcanescu, G. Collin, H. Kojima, I. Tanaka, and L. Fruchter, Phys. Rev. B {\bf50}, 4139 (1994).	
	
\bibitem{janossy95}B. Janossy, A. de Graaf, P.H. Kes, V.N. Kopylov, and T.G. Togonidze, Physica C {\bf246}, 277 (1995).	

\bibitem{timusk}For introductory reviews and references see, e.g., T. Timusk, Phys. World \textbf{18}, 31 (2005); M. Franz, Nat. Phys. \textbf{3}, 686 (2007); B.G. Levi, Phys. Today \textbf{60}, (12), 17 (2007).

\bibitem{mosqueira09}See, e.g., J. Mosqueira, L. Cabo, and F. Vidal, Phys. Rev. B \textbf{80}, 214527 (2009), and references therein. See also, L. Cabo, J. Mosqueira, and F. Vidal, Phys. Rev. Lett. \textbf{98}, 119701 (2007).

\bibitem{gammatl2223}See, e.g., D.E. Farrell, C.M. Williams, S.A. Wolf, N.P. Bansal, V.G. Kogan, Phys. Rev. Lett. 61, 2805 (1988); J.R. Thompson, D.K. Christen, H.A. Deeds, Y.C. Kim, J. Brynestad, S.T. Sekula, and J. Budai, Phys. Rev. B \textbf{41}, 7293 (1990); O. Laborde, P. Monceau, M. Potel, J. Padiou, P. Gougeon, J.C. Levet, and H. N\"oel, Physica C \textbf{162-164}, 1619 (1989); V. Hardy, A. Maignan, C. Goupil, J. Provost, Ch. Simon, and C. Martin, Supercond. Sci. Technol. \textbf{7}, 126 (1994).

\bibitem{maignan94}A. Maignan, C. Martin. V. Hardy, Ch. Simon, M. Hervieu, and B. Raveau, Physica C {\bf219}, 407 (1994).

\bibitem{kogan88a}V.G. Kogan, Phys. Rev. B {\bf38}, 7049 (1988).

\bibitem{hao92}Z. Hao and J.R. Clem, Phys. Rev. B {\bf46}, 5853 (1992).

\bibitem{buzdin94}A. Buzdin and D. Feinberg, Physica C {\bf220}, 74 (1994).

\bibitem{blk92}L.N. Bulaevskii, M. Ledvij, and V.G. Kogan, Phys. Rev. Lett. {\bf68}, 3773 (1992).

\bibitem{kogan93}V.G. Kogan, M. Ledvij, A.Yu. Simonov, J.H. Cho, and D.C. Johnston,	Phys. Rev. Lett. {\bf70}, 1870 (1993).

\bibitem{martinez92}The effect of vortex fluctuations in tilted higly anisotropic HTSC (Bi-2212) was early studied by means of torque measurements in J.C. Mart\'{\i}nez, S.H. Brongersma, A. Koshelev, B. Ivlev, P.H. Kes, R.P. Griessen, D.G. de Groot, Z. Tarnavski, and A.A. Menovsky, Phys. Rev. Lett. \textbf{69}, 2276 (1992). As in these superconductors $|M_\parallel|$ is much smaller than $|M_\perp|$, in doing that the parallel component was neglected. However, as $|\vec\tau|=\mu_0H(M_\perp \sin\theta-M_\parallel \cos\theta)$, such an approximation may break down when $\theta\to\pi/2$. 

\bibitem{kogan88}V.G.	Kogan, M.M. Fang, and S. Mitra, Phys. Rev. B {\bf38}, 11958 (1988).

\bibitem{tuominen90}M. Tuominen, A.M. Goldman, Y.Z. Chang, and P.Z. Jiang, Phys. Rev. B {\bf42}, 412 (1990).

\bibitem{pugnat95}T.P. Pugnat, G. Fillion, H. Noel, M. Ingold, and B. Barbara, Europhys. Lett. {\bf29}, 425 (1995).

\bibitem{bruckental06}Y. Bruckental, A. Shaulov, and Y. Yeshurun, Phys. Rev. B {\bf73}, 214504 (2006).

\bibitem{mosqueira07}J. Mosqueira, L. Cabo, and F. Vidal, Phys. Rev. B {\bf76}, 064521 (2007).
	
\bibitem{hao}Z. Hao, Phys. Rev. B {\bf48}, 16822 (1993).
	
\bibitem{kes91}P.H. Kes, C.J. van der Beek, M.P. Maley, M.E. McHenry, D.A. Huse, M.J.V. Menken, and A.A. Menovsky, Phys. Rev. Lett.{\bf67}, 2383 (1991).

\bibitem{klemm80}R.A. Klemm and J.R. Clem, Phys. Rev. B {\bf21}, 1868 (1980).

\bibitem{blatter92}G. Blatter, V.B. Geshkenbein, and A.I. Larkin, Phys. Rev. Lett. {\bf68}, 875 (1992).

\bibitem{unpub}The small discrepancy observed at intermediate angles, mainly in the 100 K measurement, is probably due to the crudeness of the approximation in Eq.~(\ref{blktheta}), and must be the subject of further studies. It is worth mentioning that a possible breakdown of Eq.~(\ref{blktheta}) may have implications on the interpretation of recent torque measurements in terms of the perpendicular component of the magnetization. See e.g., Ref.~15 and references therein. 

\end{references}
\end{document}